\documentclass[12pt]{article}
\usepackage{epsf,amsfonts,amssymb,epsfig,amsmath}
\renewcommand{\text}[1]{#1}

\newcommand{\be}{\begin{equation}}
\newcommand{\ee}{\end{equation}}
\newcommand{\ben}{\begin{displaymath}}
\newcommand{\een}{\end{displaymath}}
\newcommand{\bea}{\begin{eqnarray}}
\newcommand{\eea}{\end{eqnarray}}
\newcommand{\bean}{\begin{eqnarray*}}
\newcommand{\eean}{\end{eqnarray*}}
\newcommand{\nn}{\nonumber \\}
\newcommand{\ba}{\begin{array}}
\newcommand{\ea}{\end{array}}
\newcommand{\bi}{\begin{itemize}}
\newcommand{\ei}{\end{itemize}}

\newcommand{\reef}[1]{(\ref{#1})}

\def\r{\rho}

\def\l{\lambda}

\def\G{\Gamma}

\def\e{\epsilon}

\def\otaula{\begin{tabular}}
\def\ctaula{\end{tabular}}

\def\bnum{\begin{enumerate}}
\def\enum{\end{enumerate}}

\def\CR{\mathcal{R}}
\def\CM{\mathcal{M}}

\def\8M{$\CM_8$}

\def\be{\begin{equation}}
\def\ee{\end{equation}}
\def\G{\Gamma}

\def\ei{e^{\underline{i}}}

\def\e1{e^{\underline{1}}}
\def\1u{\underline{1}}
\def\2u{\underline{2}}

\def\0u{\underline{0}}
\def\e{\epsilon}
\def\target{$\CR^{1,1}\times \mathcal{M}_8$ }
\def\target2{$\CR^{1,1}\times \mathcal{M}_8$,}
\def\9G{\G_{\underline{9}}}

\newcommand{\de}{\partial}

\DeclareMathOperator{\re}{Re}
\DeclareMathOperator{\im}{Im}

\def\1f{f_1^{1/2}}
\def\2f{f_2^{1/2}}
\def\4f{f_4^{1/2}}

\newcommand{\ii}{\mathrm{i}}

\newcommand{\ft}[2]{{\textstyle\frac{#1}{#2}}}

\begin{document}

\makeatletter
\renewcommand{\theequation}{\thesection.\arabic{equation}}
\@addtoreset{equation}{section}
\makeatother

\baselineskip 18pt

\begin{titlepage}

\vfill

\begin{flushright}
Imperial/TP/2007/JG/02\\
%hep-th/yymmnnn\\
\end{flushright}

\vfill

\begin{center}
   \baselineskip=16pt
   {\Large\bf  Consistent Kaluza-Klein Reductions for General Supersymmetric
   $AdS$ Solutions}
   \vskip 2cm
      Jerome P. Gauntlett and Oscar Varela
   \vskip .6cm
      \begin{small}
      \textit{Theoretical Physics Group, Blackett Laboratory, \\
        Imperial College London, London SW7 2AZ, U.K.}
        %E-mail: j.gauntlett, d.waldram@imperial.ac.uk}
        \end{small}\\*[.6cm]
      \begin{small}
      \textit{The Institute for Mathematical Sciences, \\
        Imperial College London, London SW7 2PE, U.K.}
        %E-mail: j.gauntlett, d.waldram@imperial.ac.uk}
        \end{small}
   \end{center}

\vfill

\begin{center}
\textbf{Abstract}
\end{center}
%We provide additional evidence for the conjecture that for any
%supersymmetric solution of D=10 or D=11 supergravity consisting of a warped product of
%anti-de-Sitter space with a Riemanian manifold $M$, there is a consistent Kaluza-Klein reduction on
%$M$ to a gauged supergravity theory in $d+1$-dimensions for which the fields are dual to those in the superconformal current multiplet of
%the $d$-dimensional dual SCFT. This means that any solution of the gauged supergravity theory can be uplifted on $M$ to obtain
%an exact solution of D=10 or D=11 supergravity. We verify this conjecture for the
%most general class of supersymmetric $AdS_5$ solutions of type IIB supergravity and for
%two different classes of supersymmetric $AdS_4$ solutions of $D=11$ supergravity by constructing
%an explicit KK reduction ansatz.

For the most general supersymmetric solutions of type IIB supergravity
consisting of a warped product of $AdS_5$ with a five-dimensional manifold
$M_5$, we construct an explicit consistent Kaluza-Klein reduction on $M_5$
to minimal $D=5$ gauged supergravity. Thus, any
solution of the gauged supergravity
can be uplifted on $M_5$ to obtain
an exact solution of type IIB supergravity. We also show that
for general $AdS_4\times SE_7$ solutions, where $SE_7$ is a seven-dimensional Sasaki-Einstein manifold, and for a general class of supersymmetric solutions
that are a warped product of
$AdS_4$ with a seven-dimensional manifold $N_7$, there is an analogous
consistent reduction to minimal $D=4$ gauged supergravity.

\begin{quote}

\end{quote}

\vfill

\end{titlepage}

\setcounter{equation}{0}

%%%%%%%%%%%%%%%%%%%%%%%%%%%%%%%%%%%%%%%%%%%%%%%%%%%%%%%%%%%%%%%%%%%%%%%
%\tableofcontents
%%%%%%%%%%%%%%%%%%%%%%%%%%%%%

\section{Introduction}

A powerful method to construct solutions of $D=10$ or $D=11$
supergravity is to uplift solutions of simpler theories in
lower-dimensions. For this to work it is necessary that there is an
appropriate {\it consistent} Kaluza-Klein (KK) reduction on some
internal manifold $M$ from $D=10$ or $D=11$ down to the
lower-dimensional theory. In general, a KK expansion on $M$ leads to
a lower dimensional theory involving an infinite tower of fields.
Splitting these fields into a finite number of ``light'' fields and
an infinite tower of ``heavy'' fields\footnote{In general there is
not a sharp separation of energy scales, and hence the quotation
marks.}, the KK reduction is called consistent if it is in fact
consistent to set all of the heavy fields to zero in the equations
of motion, leaving equations of motion for the light fields only.
Clearly this is only possible if the on-shell light fields do not
source the heavy fields.

KK reductions on a circle or more generally on an $n$-dimensional torus are always consistent.
The heavy fields, which arise from modes with non-trivial dependence on
the coordinates of the torus, are all charged under the $U(1)^n$ gauge symmetry,
while the light fields, which in this case are actually massless fields, are independent of
these coordinates and hence uncharged under the gauge symmetry.
As a consequence, the heavy fields can never be sourced by the light fields alone and so the
truncation to the light fields is consistent.
Since this argument also extends to fermions, one concludes that a KK reduction of a higher-dimensional
supergravity theory on a torus can always
be consistently truncated to a lower-dimensional supergravity theory. Moreover, solutions of the lower
dimensional supergravity theory that preserve supersymmetry will uplift to supersymmetric solutions of
the higher dimensional supergravity theory.

More generally, however, consistent KK reductions are very much the
exception rather than the rule. For example, it is only in very
special circumstances that there is a consistent KK reduction on a
sphere (for further discussion see \cite{Cvetic:2000dm}). An
interesting class of examples are those associated with the
maximally supersymmetric solutions of $D=10$ and $D=11$ supergravity
that consist of products of $AdS$ spaces and spheres. Corresponding
to the $AdS_4\times S^7$ and $AdS_7\times S^4$ solutions of $D=11$
supergravity, there are consistent KK reductions on $S^7$
\cite{deWit:1986iy} and $S^4$ \cite{Nastase:1999cb,Nastase:1999kf}
to $D=4$ $SO(8)$ gauged supergravity and $D=7$ $SO(5)$ gauged
supergravity, respectively. Similarly, starting with the
$AdS_5\times S^5$ solution of type IIB supergravity there is
expected to be a consistent KK reduction to $SO(6)$ gauged
supergravity: various additional truncations were shown to be
consistent in  \cite{Cvetic:1999xp,Lu:1999bw,Cvetic:2000nc} and an
ansatz for the full metric was constructed in \cite{Khavaev:1998fb}.

We would like to view these examples as special cases of the following conjecture:
\begin{quote}
   \textsl{For any supersymmetric solution of $D=10$ or $D=11$ supergravity
that consists of a warped product of $d+1$ dimensional
anti-de-Sitter space with a Riemannian manifold $M$,
$AdS_{d+1}\times_w M$, there is a consistent Kaluza-Klein truncation
on $M$ to a gauged supergravity theory in $d+1$-dimensions for which
the fields are dual to those in the superconformal current multiplet
of the $d$-dimensional dual SCFT.}
\end{quote}
Equivalently, one can characterise the fields of the gauged
supergravity as those that contain the $d+1$-dimensional graviton
and fill out an irreducible representation of the superisometry
algebra of the $D=10$ or $D=11$ supergravity solution.
This conjecture is essentially a restricted version of one that appeared
long ago in \cite{Duff:1985jd}, for which general arguments supporting it
were put forward in \cite{sp}.

For example the $AdS_5\times S^5$ solution of type IIB, which has
superisometry algebra $SU(2,2|4)$, is dual to $N=4$ superYang-Mills
theory in $d=4$. The superconformal current multiplet of the latter
theory includes the energy momentum tensor, $SO(6)$ R-symmetry
currents, along with scalars and fermions. These are dual to the
metric, $SO(6)$ gauge fields along with scalar and fermion fields,
and are precisely the fields of the maximally supersymmetric $SO(6)$
gauged supergravity in five-dimensions.

As we have phrased the conjecture above, it is natural to try and
prove the conjecture directly from the SCFT point of view. For the
case of $AdS_3$ solutions, an argument has been made by
\cite{David:2007ak,sen}, but this needs to be modified for higher
dimension $AdS$ solutions. While we think that this is an
interesting avenue to pursue, in this paper we will verify the
conjecture for a number of cases by constructing an explicit
consistent KK reduction ansatz. By this we mean an explicit ansatz
for the higher-dimensional fields that is built from the fields of
the lower-dimensional theory with the property that it solves the
equations of motion of the higher-dimensional theory provided that
the equations of the lower-dimensional theory are satisfied. This
approach has the advantage that it allows one to uplift an explicit
solution of the lower-dimensional gauged supergravity to obtain an
explicit solution\footnote{Note that since the uplifting formulae
are local, in general, even if the lower-dimensional solution is
free from singularities one still needs to check that the higher
dimensional solution is also.} of $D=10$ or $D=11$ supergravity.

Often, for simplicity, such explicit KK reduction ans\"atze are
constructed for the bosonic fields only. This is thought to provide
very strong evidence that the ansatz can be extended to the
fermionic fields also. In fact an argument was constructed in
\cite{Cvetic:2000dm}, based on \cite{sp}, which shows that if a
consistent KK reduction has been constructed for the bosonic fields,
then the supersymmetry of the higher dimensional theory will
guarantee that the reduction can be consistently extended to the
fermionic sector. In any event, a bosonic KK ansatz certainly allows
one to uplift bosonic solutions which is the most interesting class
of solutions. One can go further and construct an ansatz for the
fermion fields and demand that the supersymmetry variation of a
bosonic configuration in higher-dimensions leads to the correct
supersymmetry variation of the bosonic configuration in
lower-dimensions. This explicitly demonstrates that a supersymmetric
bosonic solution of the lower-dimensional theory will uplift to a
supersymmetric solution of $D=10$ or $D=11$ supergravity which will
preserve at least the same amount of supersymmetry as in the
lower-dimensional theory.

In this paper we will verify the conjecture for a general class of
$AdS_5$ solutions which are dual to $N=1$ SCFTs in $d=4$ dimensions.
For this case, the bosonic fields in the superconformal current
multiplet are the energy momentum tensor and the abelian R-symmetry
current. Thus we seek a consistent truncation to minimal $D=5$
gauged supergravity whose bosonic fields are the metric (dual to the
energy momentum tensor of the SCFT) and an abelian gauge field (dual
to the R-symmetry current). For the special class of solutions of
type IIB of the form $AdS_5\times SE_5$, where $SE_5$ is a
five-dimensional Sasaki-Einstein manifold, and only the self-dual
five-form is non-vanishing, a consistent KK reduction was
constructed in \cite{Buchel:2006gb} (see also \cite{Tsikas:1986rx}).
Here we will extend this result by showing that for the most general
$AdS_{5}\times_w M_5$ supersymmetric solution of type IIB
supergravity with all of the fluxes active, that were analysed in
\cite{Gauntlett:2005ww}, the KK reduction is also consistent. We
will construct a KK ansatz for the bosonic fields and we will also
verify the consistency of the supersymmetry variations. The
analogous result for the most general supersymmetric solutions of
$D=11$ supergravity of the form $AdS_5\times_w M_6$ with
non-vanishing four-form flux \cite{GMSW}, was shown in
\cite{Gauntlett:2006ai}. Given that any $AdS_5$ solution of type IIA
supergravity can be considered to be a special case of one in
$D=11$, if we are assume that there are no $AdS_5$ solutions in type
I supergravity, the results here combined with
\cite{Buchel:2006gb,Gauntlett:2006ai} covers all $AdS_5$ solutions
in $D=10$ and $D=11$ dimensions.

We will also prove similar results for two classes of $AdS_4$
solutions of $D=11$ supergravity, both of which are dual to $N=2$
SCFTs in $d=3$. The first, and the simplest, is the Freund-Rubin
class of solutions which take the form $AdS_4\times SE_7$ where
$SE_7$ is a seven-dimensional Sasaki-Einstein manifold and the
four-form flux is proportional to the volume form of the $AdS_4$
factor. A discussion of this case appears in \cite{Duff:1984hn}.
Furthermore, our analysis is a simple extension of the analysis in
\cite{Pope:1985jg} which considered the seven-sphere viewed as a
$U(1)$ fibration over $CP^3$. The second is the class of
$AdS_4\times_w N_7$ solutions, corresponding to M5-branes wrapping
SLAG 3-cycles, that were classified in \cite{GMMW1}. It is very
plausible that this class of solutions are the most general class of
solutions with this amount of supersymmetry and with purely magnetic
four-form flux. In both cases we show that there is a consistent KK
reduction on the $SE_7$ or the $N_7$ to minimal gauged supergravity
in four spacetime dimensions. The bosonic fields of the latter
theory again consist of a metric and a $U(1)$ gauge field which are
dual to the bosonic fields in the superconformal current multiplet.
For these examples, we will be content to present the KK ansatz for
the bosonic fields only.

The general classes of supersymmetric solutions that we consider have been
analysed using $G$-structure techniques \cite{Gauntlett:2002sc,Gauntlett:2002fz}. In particular, the $G$-structure
can be characterised in terms of bi-linears constructed from the Killing
spinors. Since the results we obtain only assume supersymmetry and
$AdS$ factors one might expect that the explicit KK reduction ansatz
involves these bi-linears, and this is indeed the case.
In fact it might be illuminating to recast the known consistent KK truncations on spheres
in terms of this language, but we shall not investigate that here.

The plan of the rest of the paper is as follows. We begin in
sections 2 and 3 by considering the $AdS_4$ solutions of $D=11$
supergravity. In section 4 we consider the general class of $AdS_5$
solutions of type IIB supergravity. Only for the latter class we
will present details of our calculations and these can be found in
the appendices. In section 5 we briefly conclude.

\section{Reduction of $D=11$ supergravity on $SE_7$}
Our starting point in this section is the class of supersymmetric
solutions of $D=11$ supergravity of the form $AdS_4\times SE_7$
where $SE_7$ is a Sasaki-Einstein 7-manifold: \bea
ds^2_{11}&=&\ft{1}{4}ds^2(AdS_4)+ds^2(SE_7)\nn G&=&\ft{3}{8}
\textrm{vol}(AdS_4) . \eea Here $\textrm{vol}(AdS_4)$ is the volume
4-form of the unit radius $AdS_4$ metric $ds^2(AdS_4)$ and we have
normalised the Sasaki-Einstein metric $ds^2(SE_7)$ so that
$Ric(SE_7)=6g(SE_7)$ (the same as for the unit radius metric on the
round seven-sphere). The Sasaki-Einstein metric has a Killing vector
which is dual to the R-symmetry of the dual $N=2$ SCFT in $d=3$.
Introducing coordinates so that this Killing vector is
$\partial_\psi$, locally, the Sasaki-Einstein metric can be written
\be ds^2(SE_7)=(d\psi+\sigma)^2+ds^2(M_6) \ee where $ds^2(M_6)$ is
locally K\"ahler-Einstein with K\"ahler form $J$, normalised so that
$Ric(M_6)=8 g(M_6)$ and $d\sigma=2J$.

We now construct an ansatz which leads to a consistent truncation,
at the level of bosonic fields, to gauged supergravity in $D=4$.
Specifically, we consider \bea
ds^2_{11}&=&\ft{1}{4}ds^2_4+(d\psi+\sigma+
\ft{1}{4}A)^2+ds^2(M_6)\nn G&=&\ft{3}{8} \textrm{vol}_4- \ft{1}{4}
*_4  F_2 \wedge J \eea where $ds^2_4$ is an arbitrary metric on
a four-dimensional spacetime, $\textrm{vol}_4$ is its associated
volume form, and $A$ and $F_2=dA$ are one- and two-forms on this
spacetime with a normalisation chosen for convenience. Substituting
this into the $D=11$ equations of motion \cite{Cremmer:1978km} (we
use the conventions of \cite{Gauntlett:2002fz}), \bea\label{d=11eq}
R_{AB}-\frac{1}{12}(G_{A C_1C_2C_3}G{_{B}}{^{C_1C_2C_3}}-
\frac{1}{12}g_{AB}G^2)&=&0\nn d*_{11}G+\frac{1}{2}G\wedge G&=&0\nn
dG&=&0 \eea we find that the metric $g_{\mu\nu}$, corresponding to
$ds^2_4$, and $F_2$ must satisfy \bea \label{eomD=4}
R_{\mu\nu}=-3g_{\mu\nu}+\ft12F_{\mu\rho}F_{\nu}{}^\rho
-\ft{1}{8}g_{\mu\nu}F_{\rho\sigma}F^{\rho\sigma} \nn d*_4F_2=0 .
\eea These are precisely the equations of motion of minimal gauged
supergravity in $D=4$ \cite{Fradkin:1976xz,Freedman:1976aw}.

Thus we have shown the consistency of the KK reduction at the level
of the bosonic fields. In particular, any solution of the minimal
gauged supergravity, which were systematically studied in
\cite{Caldarelli:2003pb}, can be uplifted on an arbitrary
seven-dimensional Sasaki-Einstein manifold to a solution of $D=11$
supergravity

\section{Reduction of $D=11$ supergravity on a SLAG-3 Flux Geometry}

Let us now consider the general class of supersymmetric warped
product solutions of the form $AdS_4 \times_w {\cal N}_7$  with
purely magnetic four-form flux which are dual to $N=2$ SCFTs in
$d=3$ \cite{GMMW1}. We call these geometries SLAG-3 flux geometries,
since they can be derived from a class of geometries that correspond
to M5-branes wrapping special lagrangian (SLAG) three-cycles in a
$SU(3)$ holonomy manifold - for further details see \cite{GMMW1}. It
is quite possible that this class of geometries is the most general
class of $AdS_4$ geometries with this amount of supersymmetry and
with purely magnetic four-form flux, but this has not been proven.

%Since the dual SCFT has $N=2$ supersymmetry and a $U(1)$ R-symmetry, we again seek a consistent
%truncation from $D=11$ down to $D=4$, $N=2$ gauged supergravity.

The $D=11$ metric of the SLAG-3 flux geometry is given by \be
ds^2_{11} =\lambda^{-1} ds^2(AdS_4) +   d s^2(\mathcal{N}_7) \ee
where $ds^2(AdS_4)$ has unit radius and the warp factor $\lambda$ is
independent of the coordinates of $AdS_4$. $\mathcal{N}_7$ has a
local $SU(2)$ structure which is specified by three one-forms and
three self-dual two-forms $J^1,J^2,J^3$. One of the one-forms is
dual to a Killing vector that also preserves the flux: this is dual
to the R-symmetry of the corresponding $N=2$ SCFT. Introducing local
coordinates so that this Killing vector is given by $\partial_\phi$
we have
\begin{equation}
   d s^2(\mathcal{N}_7) =
      d s^2(\mathcal{M}_{SU(2)})
      + w \otimes w
      + \frac{\lambda^2d\rho^2}{4(1-\lambda^3\rho^2)}
      + \frac{\lambda^2\rho^2}{4}d\phi^2 ,
\end{equation}
$\mathcal{M}_{SU(2)}$ is a four-dimensional space where the $J^a$
live. The three one-forms mentioned above are $w$,
$(\lambda/2{\sqrt{1-\lambda^3\rho^2} })d\rho$ and
$(\lambda\rho/2)d\phi$. In addition we must have
\bea \label{101} d[\l^{-1}\sqrt{1-\l^3\r^2} w]&=&
   \l^{-1/2}J^1+ \frac{\lambda^2 \rho}{2\sqrt{1-\lambda^3\rho^2} }  w \wedge d\rho ,\nn
%m^{-1}d[\l^{-1}\sqrt{1-\l^3\r^2}e^5]-\l^{-1/2}J^1-\l\r e^5\wedge
%e^6&=&0,\\
d\left(\l^{-3/2}J^3\wedge w - \frac{\lambda
\rho}{2\sqrt{1-\lambda^3\rho^2} }  J^2 \wedge d\rho \right)
   &=&0,\nn
d\left(J^2\wedge w + \frac{1}{2 \lambda^{1/2} \rho
\sqrt{1-\lambda^3\rho^2}} J^3\wedge d\rho \right)
   &=&0 .
%\\\label{101111}
%\l^2[-d[\l^{-2}\sqrt{1-\l^3\r^2}J^2]+3m(\l^{-3/2}J^3\wedge e^5-\r
%J^2\wedge e^6)]&=&\star_7F.
\eea
Finally the 4-form flux is given by
\bea\label{515}
G=d\phi\wedge
   d\left(\frac{1}{2}\l^{-1/2}\sqrt{1-\l^3\r^2}J^3\right) .
\eea
An explicit example of a solution to these equations was given in~\cite{gkw}
as discussed in \cite{GMMW1}.

We now consider the KK reduction ansatz:
 \bea ds^2_{11}&=& \lambda^{-1} ds^2_4+ ds^2(\hat {\cal N}_7)
\nn G &=& \hat G + F_2 \wedge Y +*_4F_2 \wedge X \label{KKansatz}
\eea
where $ds^2_4$ is a line element and $F_2=dA$ is a two-form on a
four-dimensional spacetime. In addition $ds^2(\hat {\cal N}_7)$ is
the expected deformation of $ds^2({\cal N}_7)$, given by
\begin{equation}
   d s^2(\hat {\mathcal{N}}_7) =
      d s^2(\mathcal{M}_{SU(2)})
      + {w}\otimes {w}
      + \frac{\lambda^2d\rho^2}{4(1-\lambda^3\rho^2)}
      + \frac{\lambda^2\rho^2}{4}(d\phi+A)^2 ,
\end{equation}
$\hat G$ is the expected deformation of the four-form flux appearing in \reef{515}
\bea
\hat G=(d\phi+A)\wedge
   d\left(\frac{1}{2}\l^{-1/2}\sqrt{1-\l^3\r^2}J^3\right)
\eea
and the two-forms $X$ and $Y$ are given by \bea && \label{alpha2} X
= - \frac{1}{2} (\lambda^{-1/2} J^1 + \frac{\lambda^2 \rho}{2\sqrt{1-\l^3\rho^2}} \ {\omega}
\wedge d{\rho} )\nn && \label{beta2} Y = - \frac{1}{2}
\lambda^{-1/2} \sqrt{1-\lambda^3 \rho^2} J^3 .
\end{eqnarray}

Substituting this ansatz into the equations of motion of $D=11$
supergravity \reef{d=11eq} and using \reef{101} we find that all
equations are satisfied provided that the equations of motion
(\ref{eomD=4}) of minimal gauged supergravity in $D=4$ are
satisfied. This again shows the consistency of the truncation, at
the level of the bosonic fields.

\section{Reduction of IIB on general $M_5$}

We now turn to the general class of supersymmetric $AdS_5 \times_w
M_5$ solutions of IIB supergravity with all fluxes active that were
analysed in \cite{Gauntlett:2005ww}. Such solutions are dual to
$N=1$ SCFTs in $d=4$ which all have a $U(1)$ $R$-symmetry. We will
show that there is a consistent KK reduction on $M_5$ to minimal gauged
supergravity in $D=5$. This case is more involved than the previous
two and so we have included some details of the calculation in the appendices.

\subsection{Internal geometry and fluxes}

%The most general minimally supersymmetric bosonic solutions of IIB
%supergravity containing an $AdS_5$ factor in the metric and with all
%IIB fluxes turned on, in a way consistent with the symmetries of
%$AdS_5$, were analysed in \cite{Gauntlett:2005ww}. Specifically,
We begin by summarising the results of \cite{Gauntlett:2005ww}. The
ten-dimensional metric is a warped product of $AdS_5$ with a
five-dimensional Riemannian manifold $M_5$,
\begin{equation} \label{metansatz}
ds^2_{10} = e^{2\Delta} \left[ ds^2(AdS_5)+ ds^2(M_5) \right] \; ,
\end{equation}
where the warp factor $\Delta$ is a real function on $M_5$.
All fluxes are active:
%We call
%$f_5$, $g_3$, $p_1$, $q_1$ the IIB forms corresponding to this
%configuration $AdS_5 \times_w M_5$.
in order to preserve the spatial $SO(4,2)$ isometry, the one-forms $P$, $Q$
and the complex three-form $G$ lie entirely on
the internal $M_5$, and the five-form is taken to be
\begin{equation} \label{genansatz}
F= f \left( \textrm{vol}_{AdS_5} + \textrm{vol}_{M_5} \right) \; ,
\end{equation}
where $f$ is a constant and $\textrm{vol}$ is the volume form
corresponding to each of the metrics in the r.h.s.~of
(\ref{metansatz}). We use the same conventions as in \cite{Gauntlett:2005ww} and
some of this is recorded in appendix A.

The manifold $M_5$ is equipped with two spinors $\xi_1$,
$\xi_2$ of Spin(5) subject to a set of differential and algebraic
constraints
%(equations (\ref{sfive})--(\ref{stwo}) of Appendix
%\ref{ppKKansbos})
arising from the IIB Killing spinor equations. The spinors $\xi_1$,
$\xi_2$ define a local identity structure on $M_5$, which can be
conveniently characterised in terms of a set of forms, bi-linear in
$\xi_1$, $\xi_2$, consisting of a real scalar $\sin \zeta$, a
complex scalar $S$, a real one-form $K_5$, and two complex one-forms
$K, K_3$. These satisfy the following differential conditions
\begin{align}
   e^{-4\Delta} d (e^{4\Delta} S) &= 3 i  K  \nn
   e^{-6\Delta} D(e^{6\Delta} K_3)
      &= P\wedge K_3^* - 4 i  W - e^{-2\Delta}*G \nn
   e^{-8\Delta} d (e^{8\Delta} K_5)
      &= 4 \sin\zeta V - 6  U \label{K5-summ}
\end{align}
where $D(e^{6\Delta} K_3) \equiv d(e^{6\Delta} K_3 ) -i Q \wedge
e^{6\Delta} K_3$. In (\ref{K5-summ}),  $U, V$ are real two-forms and
$W$ is a complex two-form that can be constructed as bi-linears in
$\xi$ and moreover can be expressed in terms of the identity
structure:
\begin{align}
\label{UVWexp}
   U &= \frac{1}{2(\cos^2\zeta-|S|^2)}\big(
       \ii \sin\zeta K_3 \wedge K_3^*
       + \ii K \wedge K^* - 2\im S^*K\wedge K_5\big) ,\nn
   V &= \frac{1}{2\sin\zeta(\cos^2\zeta-|S|^2)}\big(
       \ii \sin\zeta K_3\wedge K_3^*
       \nn & \qquad\qquad\qquad  {}
       + \ii [\sin^2\zeta+|S|^2] K\wedge K^*
       - 2\im S^*K\wedge K_5\big) ,\nn
   W & = \frac{1}{\sin\zeta(\cos^2\zeta-|S|^2)}\big(
       \cos^2\zeta K_5
       + \re S^*K
       + \ii\sin\zeta\im S^*K\big)\wedge K_3 .
\end{align}

In addition, one also has the algebraic constraint
\begin{equation}
\label{holo-sum}
   \ii_{K_3^*} P = 2\,\ii_{K_3} d\Delta \ ,
\end{equation}
the five-form flux is given by (\ref{genansatz}) with
\begin{equation}
\label{f-summ}
   f = 4 e^{4\Delta}\sin\zeta ,
\end{equation}
the three-form flux is given by
\begin{equation}
\label{flux-summ}
\begin{aligned}
   \left(\cos^2\zeta-|S|^2\right) &\, e^{-2\Delta} *G
      \\ &
      = 2P \wedge K_3^*
      - \left(4d\Delta + 4i  K_4
         - 4 i \sin\zeta K_5\right) \wedge K_3
      \\ & %\qquad
      + 2 * \left( P\wedge K_3^*\wedge K_5
         - 2d\Delta\wedge K_3\wedge K_5 \right)~,
\end{aligned}
\end{equation}
where $\sin \zeta K_4 = K_5 + \re ( S^*K )$,  and the metric can be
written
\begin{equation}
\begin{aligned}
d s^2(M_5) & = \frac{(K_5)^{2}}{\sin^2\zeta+|S|^2}
    + \frac{K_3\otimes K_3^*}{\cos^2\zeta-|S|^2}
    + \frac{|S|^2}{\cos^2\zeta-|S|^2}
        \left(\im{S^{-1}K}\right)^{2}
        \\ & \qquad
    + \frac{|S|^2}{\sin^2\zeta}\;
        \frac{\sin^2\zeta+|S|^2}{\cos^2\zeta-|S|^2}
        \left(\re{S^{-1}K}
            + \frac{1}{\sin^2\zeta+|S|^2}K_5\right)^{2}~.
%\frac{1}{N^2}(K_5)^2+\frac{|S|^2}{1-N^2}\rm{Im}\left(\frac{1}{S}
%K\right)^2\\ &+\frac{|S|^2}{\sin^2\zeta
%N^2(1-N^2)}\rm{Re}\left(\frac{N^2}{S}K+K_5\right)^2
%+\frac{1}{1-N^2}\left(K_3\otimes K_3^*\right)^2~.
\label{undefmetM5}
\end{aligned}
\end{equation}

Finally, the vector dual to $K_5$ is a Killing vector of the metric
(\ref{undefmetM5}) that also generates a symmetry of the full
solution: $\mathcal{L}_{K_5}\Delta=i_{K_5}P=\mathcal{L}_{K_5}G=0$.
The above constraints arising from supersymmetry ensure that all
equations of motion and Bianchi identities are satisfied.

\subsection{KK reduction}

We now construct the ansatz for a KK reduction from type IIB on the
general $M_5$ that we discussed in the last subsection. We shall
show that there is a consistent reduction to minimal $D=5$ gauged
supergravity.

On $M_5$ the vector field dual to the one-form $K_5$ is Killing and
corresponds to the R-symmetry in the $d=4$ dual SCFT. If one
introduces coordinates such that this dual vector field is
$3\partial_\psi$, we would like to shift $d\psi$ by the gauge field
$A$: noting that $||K_5||^2=(\sin^2\zeta+|S|^2)$ this means that we
should make the shift
\begin{equation} \label{shiftU1}
K_5 \quad \longrightarrow \quad \hat K_5 = K_5 + (\sin^2\zeta+|S|^2)\frac{A}{3} \; .
\end{equation}
In particular, given (\ref{metansatz}), our ansatz for the $D=10$ type IIB metric is then
\begin{equation} \label{KKmetric}
ds^2_{10} = e^{2\Delta} \left[ ds^2_5 + ds^2(\hat M_5) \right] \;
\end{equation}
where $ds^2_5$ is an arbitrary
metric on five-dimensional spacetime, and $ds^2(\hat M_5)$ is the metric $ds^2(M_5)$
in (\ref{undefmetM5}) after the shift \reef{shiftU1}.
%\begin{equation} \label{shiftU1}
%K_5 \quad \longrightarrow \quad \hat K_5 = K_5 + A \; .
%\end{equation}

The KK ansatz for the five-form and the complex three-form of type
IIB reads:
\begin{eqnarray}
&& F_5 = \hat F_5 + F_2 \wedge \ft{1}{3} e^{4\Delta} \hat *_5 V  +
*_5 F_2 \wedge \ft{1}{3} e^{4\Delta}  V  \nn
&& \label{KKansatzG} G = \hat G+ F_2 \wedge  \ft{1}{3} e^{2\Delta} K_3
%&& \label{KKansatzP}  P= \hat p \\
%&& \label{KKansatzQ} Q= \hat q \label{AKKansatz}
\end{eqnarray}
where $F_2=dA$, $\hat F_5$ and $\hat G$ are the five-form and
three-form flux of the undeformed solution on $M_5$ after we make
the shift (\ref{shiftU1}), $V$, $K_3$ are the bi-linears on $M_5$
introduced in the previous subsection\footnote{The bi-linear $V$ is
not affected by the shift (\ref{shiftU1}): choosing the convenient
frame of Appendix B of \cite{Gauntlett:2005ww} one can check that
all $K_5$ dependence of $V$ in equation (\ref{UVWexp}) drops out.},
and $\hat *_5$ and $*_5$ are, respectively, the Hodge duals with
respect to the metrics $ds^2(\hat M_5)$ and $ds^2_5$ in
(\ref{KKmetric}). Notice that Since the one-forms $P$ and $Q$ of the
undeformed solution on $M_5$ are independent of $K_5$, they remain
the same as they were.

In appendix \ref{AppKKansbos} we provide some details of how we constructed this particular ansatz.
In particular, a long calculation shows that the ansatz (\ref{KKmetric}), (\ref{KKansatzG}) with $P,Q$ unchanged
satisfies all of the IIB equations of motion and Bianchi identities, provided that $ds^2_5$ and $F_2$
satisfy
\begin{eqnarray}
&& \label{D=5Einstein} R_{\mu \nu} = -4 g_{\mu \nu} +\ft{1}{6}
F_{\mu \lambda} F_\nu{}^\lambda -\ft{1}{36}  g_{\mu \nu} F_{\lambda
\rho} F^{\lambda \rho}  \\ && \label{D=5eomF} d*_5 F_2 - \ft{1}{3}
F_2 \wedge F_2 =0 .
\end{eqnarray}
These are precisely the equations of motion of minimal $D=5$ gauged
supergravity \cite{Gunaydin:1983bi}. This shows the consistency of
the truncation of the bosonic sector.

The truncation is, moreover, consistent at the level of the
variations of the IIB fermion fields (see Appendix \ref{AppKKansfer}
for the details). On the one hand we find that the supersymmetry
variations of the dilatino $\lambda$ and of the internal
components of the gravitino $\Psi_M$ identically vanish.
%\begin{eqnarray}
%&& \delta \lambda =0 \\
%&& \delta \Psi_a =0 .
%\end{eqnarray}
On the other hand, the external components of the IIB gravitino
variation reduce to
\begin{equation}
\delta \psi_\alpha = D_\alpha \varepsilon -\ft12 \rho_\alpha
\varepsilon + \ft{i}{2} A_\alpha \varepsilon + \ft{i}{24} F_{\beta
\gamma} (\rho_\alpha{}^{\beta \gamma} -4 \delta_\alpha^\beta
\rho^\gamma ) \varepsilon ,
\end{equation}
where $\psi_\alpha$ is the $D=5$ gravitino and $\varepsilon$ a $D=5$
spinor. This is the gravitino variation corresponding to minimal
$D=5$ gauged supergravity.

To summarise, we have shown that any bosonic solution of $D=5$ supergravity can be
uplifted to $D=10$ using a general supersymmetric solution by means of the KK ansatz (\ref{KKmetric}),
(\ref{KKansatzG}). Moreover, if the five-dimensional bosonic solution is supersymmetric\footnote{Such solutions
were classified in \cite{Gauntlett:2003fk}.} then so will be the uplifted ten-dimensional solution.

\section{Conclusion}

In this paper we have constructed explicit consistent KK reduction
ans\"atze for general classes of $AdS_5$ solutions in type IIB
supergravity and $AdS_4$ solutions in $D=11$ supergravity. Our
results can be extended to other classes of supersymmetric solutions
that have been classified. It would be nice to show for the
$AdS_5\times_w M_6$ solutions of $D=11$ supergravity, classified in
\cite{Lin:2004nb}, which are dual to $N=2$ SCFTs in $d=4$, that
there is a consistent KK reduction to the $SU(2)\times U(1)$ gauged
supergravity of \cite{Romans:1985ps}. A similar result in type IIB
requires an analogous classification of $AdS_5\times_w M_5$
solutions that are dual to $N=2$ SCFTs in $d=4$, which has not yet
been carried out.

There are several classes of $AdS_4$ solutions of $D=11$
supergravity that can be considered. For example, one can consider
$AdS_4\times N_7$ solutions of $D=11$ where $N_7$ has weak $G_2$
holonomy \cite{Acharya:1998db,Morrison:1998cs} or the $AdS_4\times_w
N_7$ solutions that arise from $M5$-branes wrapping associative
3-cycles that were analysed in \cite{GMMW1}. These solutions are
dual to $N=1$ SCFTs in $d=3$, which have no $R$-symmetry, and so one
expects a consistent KK reduction on $N_7$ to a $N=1$ supergravity
whose field content is just the metric and fermions. In fact it is
easy to show that there is a consistent reduction to the $N=1$
supergravity of \cite{Townsend:1977qa}. Similarly, the $AdS_4\times
N_7$ solutions of $D=11$ where $N_7$ is tri-Sasaki
\cite{Acharya:1998db,Morrison:1998cs}, are dual to $N=3$ SCFTs in
$d=3$ and there should be a consistent KK reduction to an $SO(3)$
gauged supergravity in $D=4$. Additional $AdS_3$ and $AdS_2$
solutions of $D=11$ supergravity studied in \cite{GMMW1,Mac
Conamhna:2006nb,Figueras:2007cn} can also be considered.

The consistency of the KK truncation makes it manifest from the
gravity side that SCFTs with type a IIB or $D=11$ dual share common
sectors. For example, if we consider such SCFTs in $d=4$, the black
hole solutions of minimal gauged supergravity constructed in
\cite{Gutowski:2004ez} should be relevant for any of the SCFTs. It
would be interesting to pursue this further.

\section*{Acknowledgements}
We would like to thank Marco Caldarelli, Mike Duff, Oisin Mac
Conamhna, Daniel Freedman, Jaume Gomis, Chris Pope, Leonardo
Rastelli, Ashoke Sen, Kelly Stelle, Marika Taylor, Arkady Tseytlin,
Daniel Waldram and Toby Wiseman for helpful discussions. JPG is
supported by an EPSRC Senior Fellowship and a Royal Society Wolfson
Award. JPG would like to thank the Galileo Galilei Institute for
Theoretical Physics for hospitality. OV is supported by the Spanish
Ministry of Science and Education through a postdoctoral fellowship
and partially through the research grant FIS2005-02761.

\appendix

\section{IIB supergravity conventions} \label{AppIIB}

We quote here our conventions for IIB supergravity
\cite{Schwarz:1983qr,Howe:1983sr}, that follow those of
\cite{Gauntlett:2005ww}. The bosonic ten-dimensional fields
consist of a metric and the following set of form field stregths:
a complex one-form $P$, a complex three-form $G$
and a real five-form $F_5$, subject to the following equations of motion:
\begin{eqnarray}
&&  R_{MN} = P_M P_N^*+P_NP^*_M
      + \ft{1}{96}F_{MP_1P_2P_3P_4}F_N^{~P_1P_2P_3P_4}  \nonumber \\ && \quad \qquad
      + \ft{1}{8}\left(
         G_M{}^{P_1P_2}G^*_{NP_1P_2} + G_N{}^{P_1P_2}G^*_{MP_1P_2}
         - \ft{1}{6}g_{MN}G^{P_1P_2P_3}G^*_{P_1P_2P_3} \right) , \label{IIBEinstein} \\
&& \label{IIBeomF}  * F_5 = F_5 \; , \\
&& \label{IIBeomG}  D* G-P \wedge *G^* + i G \wedge F_5=0 \; , \\
&& \label{IIBeomP}  D* P + \ft{1}{4} G \wedge * G =0 \; .
\end{eqnarray}
We are working in the formalism where $SU(1,1)$ is realised
linearly. In particular there is a local $U(1)$ invariance and $Q_M$
acts as the corresponding gauge field. Note that $Q_M$ is a
composite gauge field with field strength given by $d Q = -\ii
P\wedge P^*$. Since $G$ has charge 1 and $P$ has charge 2 under this
$U(1)$ we have the covariant derivatives: $D*G \equiv d* G -i Q
\wedge *G$ and $D*P \equiv d*P-2iQ \wedge *P$. We also need to
impose the Bianchi identities
\begin{eqnarray}
&& \label{IIBBianchiF}  dF_5-\ft{i}{2} G \wedge G^* =0  ,\nn
&& \label{IIBBianchiG} DG+P \wedge G^* =0  , \nn
&& \label{IIBBianchiP} DP =0 .
\end{eqnarray}

The IIB fermionic fields consist of a gravitino $\Psi_M$ and a
dilatino $\lambda$. For supersymmetric bosonic solutions, the
variations under supersymmetry of the fermion fields,
\begin{equation}
\delta \lambda = i \Gamma^M P_M \epsilon^c + \ft{i}{24} \Gamma^{P_1
P_2 P_3} G_{P_1 P_2 P_3} \epsilon ,  \label{susylambda}
\end{equation}
\begin{equation}
\delta \Psi_M = D_M \epsilon -\ft{1}{96} \left(\Gamma_M{}^{P_1 P_2
P_3} G_{P_1 P_2 P_3} - 9 \Gamma^{P_1 P_2} G_{M P_1 P_2} \right)
\epsilon^c + \ft{i}{192} \Gamma^{P_1 P_2 P_3 P_4} F_{ M P_1 P_2 P_3
P_4} \epsilon ,  \label{susypsi}
\end{equation}
must vanish. The spinor $\epsilon$ has composite $U(1)$ charge +1/2 so that
$D_M\e=\left(\nabla_M-\frac{\ii}{2}Q_M\right)\e$.

\section{IIB reduction: bosonic sector} \label{AppKKansbos}

We now derive the KK reduction ansatz (\ref{KKansatzG}) for the type
IIB bosonic fields Recall that the vector field dual to the
bi-linear $K_5$ is Killing and that $||K_5||^2=(\sin^2\zeta+|S|^2)$.
We therefore need to make the shift
\begin{equation}\label{dipsy}
K_5 \quad \longrightarrow \quad \hat K_5 = K_5 +
(\sin^2\zeta+|S|^2)\frac{A}{3} \;
\end{equation}
in the metric of the undeformed solution to obtain:
\begin{equation} \label{KKmetric2}
ds^2_{10} = e^{2\Delta} \left[ ds^2_5 + ds^2(\hat M_5) \right] \; .
\end{equation}
In fact for any $p$-form $\beta_p$ on $M_5$ we can define a $\hat \beta_p$ in $\hat M_5$
via
\begin{equation} \label{hattedform}
\hat \beta_p = \beta_p + \ft13 A \wedge \ii_{K_5} \beta_p \; ,
\end{equation}
where $\ii_{K_5}$ is the interior product with respect to the vector
dual to the one-form $K_5$. If we restrict to forms $\beta_p$ whose
Lie-derivative with respect to the Killing vector dual to $K_5$
vanish, it is useful in the calculations below to note that
\bea \label{dhattedform} d \hat \beta_p &=& d \beta_p  - \ft13 A
\wedge d \ii_{K_5} \beta_p +\ft13 F_2 \wedge \ii_{K_5} \beta_p\nn
&=&d \beta_p + \ft13 A \wedge \ii_{K_5} d\beta_p +\ft13 F_2 \wedge
\ii_{K_5} \beta_p \nn &\equiv&\widehat{d \beta_p} + \ft13 F_2 \wedge
\ii_{K_5} \beta_p \; . \eea

We now propose the following KK ansatz for the five-form and complex
three-form field strengths. We first take the fluxes of the
undeformed $AdS_5 \times_w M_5$ solution, and make the shift
(\ref{hattedform}) to obtain $\hat F_5$ and $\hat G$. We then
introduce a set of forms $\beta_3$, $\beta_2$, $\alpha_1$, $\alpha
_0$ on $M_5$, which we take to be invariant under the action of the
Killing vector\footnote{This is a natural condition to impose. If we
introduce coordinates so that the Killing vector field dual to $K_5$
is $3\partial_\psi$, then the condition says that the components of
the forms must be independent of $\psi$.}, and write
\begin{equation}
\begin{aligned} \label{AKKansatz}
&  F_5 = \hat F_5 + F_2 \wedge \hat  \beta_3 +*_5 F_2 \wedge \hat  \beta_2 , \\
&  G = \hat G+ F_2 \wedge \hat \alpha_1 +*_5 F_2 \hat \alpha_0 . \\
\end{aligned}
\end{equation}
The IIB forms $P$ and $Q$ in the KK ansatz are taken to be the same as those in the
undeformed solution.

For the KK reduction ansatz (\ref{KKmetric2}), (\ref{AKKansatz}) to
be consistent, it must satisfy the IIB field equations
(\ref{IIBEinstein})--(\ref{IIBBianchiP}) when the $D=5$ equations
(\ref{D=5Einstein}) and (\ref{D=5eomF}) for $ds_5^2$, $F_2$ are
satisfied. To carry out these calculations it is useful to use the
orthonormal frame $e^a$, $a=1,\ldots , 5$, on $M_5$ that was
introduced in appendix B of \cite{Gauntlett:2005ww} which, in
particular, contains
\begin{equation} \label{e1frame}
e^1= \ft{3}{h} K_5 \; , \quad h= \ft13 \sqrt{\sin^2\zeta + |S|^2} \;
.
\end{equation}
For $\hat M_5$ we use the corresponding frame obtained by the prescription \reef{dipsy}.

The requirement that the fields obey the field equations
(\ref{IIBeomF})--(\ref{IIBBianchiP}) translates into a set of
differential and algebraic equations relating the undeformed forms
$\beta_3$, $\beta_2$, $\alpha_1$, $\alpha_0$ to the undeformed
fluxes $G$, $F_5$, $P$, $Q$ and metric on $M_5$:
\begin{equation}
\begin{aligned}
 \label{Asystem}
& d\beta_2 = \ft{i}{2} (\alpha_0^* G - \alpha_0 G^*) , \\
& \ft13 \ii_{K_5} \beta_3 = -\ft13 \beta_2 +\ft{i}{2} \alpha_1 \wedge \alpha_1^* , \\
& d \beta_3 = \ft{i}{2} ( G \wedge \alpha_1^* - G^* \wedge \alpha_1) - \ft13 \ii_{K_5} F_5 , \\
& \ft13 \ii_{K_5} \beta_2 = \ft{i}{2} ( \alpha_0^* \alpha_1
- \alpha_0 \alpha_1^*) , \\
& D\alpha_1 +P \wedge \alpha_1^* + \ft13 \ii_{K_5} G =0 ,   \\
& D\alpha_0 +P  \alpha_0^* =0 , \\
& \ii_{K_5} \alpha_1 = -\alpha_0 , \\
& \beta_3= *_5 \beta_2 , \\
& \ft13 e^{4 \Delta} \ii_{K_5}  *_5 \alpha_1 = -i \alpha_1 \wedge
\beta_2
+ i \alpha_0 \beta_3 , \\
&-\ft13 e^{4 \Delta}  *_5 \alpha_1 = \ft13 e^{4\Delta} \alpha_0
\ii_{K_5}
\textrm{vol}_{M_5} + i \alpha_1 \wedge \beta_3 , \\
& D(e^{4 \Delta}  *_5 \alpha_1) -P \wedge e^{4 \Delta}
*_5
\alpha_1^* + iG\wedge \beta_2 -i \alpha_0 f\textrm{vol}_{M_5} =0 , \\
& \alpha_1 \wedge *_5 \alpha_1 = \alpha_0^2 \ \textrm{vol}_{M_5} ,
\end{aligned}
\end{equation}
where $\alpha_0, \alpha_1$ both carry charge 1 under the composite $U(1)$ gauge-field
so that e.g. $D\alpha_1 \equiv d\alpha_1 - i Q \wedge \alpha_1$.
%, $D\alpha_0
%\equiv d\alpha_0 - i Q  \alpha_0$ and $D(e^{4 \Delta}  *_5 \alpha_1)
%\equiv d(e^{4 \Delta} *_5 \alpha_1) -iQ \wedge e^{4 \Delta}  *_5
%\alpha_1$.

We must also demand that the KK ansatz satisfies the Einstein
equations. After substitution of (\ref{AKKansatz}), and imposing for
simplicity $\beta_ 3 = *_5\beta_2$ (one of the conditions in
\reef{Asystem}) we find that the external, $\mu \nu$, components of
the Einstein equation (\ref{IIBEinstein}) read
\begin{eqnarray}
&& R_{\mu\nu} = -4 g_{\mu \nu} -k_1 F_{\mu \lambda} F^\lambda{}_\nu
-k_2 g_{\mu \nu} F_{\lambda \rho} F^{\lambda \rho} \label{AEinmunu}
\end{eqnarray}
where $k_1$, $k_2$ are functions on
$M_5$ given by
%depending on the unknown forms defining the KK ansatz.
%Specifically,
%
\begin{eqnarray}
&& \label{AEin1} k_1= \ft{1}{4} \big[ e^{-8\Delta} \beta_{2  ab}
\beta_2^{ab} +2 e^{-4\Delta} \alpha_0 \alpha_0^*
+2 e^{-4\Delta} \alpha_1^a \alpha_{1 a}^* + 2h^2 \big] , \\
&& \label{AEin2} k_2= \ft{1}{16} \big[ e^{-8\Delta} \beta_{2  ab}
\beta_2^{ab} +3 e^{-4\Delta} \alpha_0 \alpha_0^* + e^{-4\Delta}
\alpha_1^a \alpha_{1 a}^*  \big] .
\end{eqnarray}
%where $\beta_{2ab}$, $\alpha_{1 a}$ are the components of the
%unknown forms in some frame.
Comparing with (\ref{D=5Einstein}) we see that we
require\footnote{The possibility that $k_1$, $k_2$ and $k_3$, below,
cannot be chosen to be constant is a potential source of
inconsistency of the KK reduction; a similar issue has been
discussed for other reductions in \cite{Duff:1984hn,Hoxha:2000jf}.}
that $k_1=1/6$ and $k_2=1/36$.

The mixed, $\mu a$, components of the Einstein equations (\ref{IIBEinstein}) give:
\begin{eqnarray}
&& \nabla^\rho F_{ \rho \mu} + \ft{k_3}{4} \epsilon_{\mu \nu \lambda
\rho \sigma} F^{\nu \lambda} F^{\rho \sigma} = 0 \label{AeomF2}
\end{eqnarray}
with
\begin{eqnarray}
&& \label{AEin3} k_3 = \ft{1}{8h} \delta^{1a} \big[ e^{-8\Delta}
\epsilon_{abcde} \beta_2^{bc} \beta_2^{de} + 4 e^{-4\Delta}
(\alpha_0 \alpha_{1  a}^* + \alpha_0^* \alpha_{1  a}) \big] .
\end{eqnarray}
%and here we are using the frame have selected a frame $e^a$ with $e^1$ given by
%(\ref{e1frame}) (hence the $\delta^{1a}$).
Comparing with (\ref{D=5eomF}) we see that we demand $k_3=-1/3$.

Finally, the internal, $ab$, components of the Einstein equations
(\ref{IIBEinstein}) give one more relation among the unknown coefficients in the KK
ansatz:
\begin{eqnarray}
 \label{AEin4} && 4 e^{-8\Delta} \beta_{2  ac} \beta_2^c{}_b + 2 e^{-4\Delta}
(\alpha_{1 a} \alpha_{1 b}^* + \alpha_{1 a}^* \alpha_{1 b})
\nonumber
\\ && + \delta_{ab} \left( e^{-8\Delta} \beta_{2  cd} \beta_2^{cd}
+ e^{-4\Delta} (\alpha_0 \alpha_0^* -  \alpha_1^c \alpha_{1  c}^*)
\right) = 4h^2 \delta_{a1} \delta_{b1} .
\end{eqnarray}
%where again a frame with $e^1$ as in (\ref{e1frame}) has been used.

After considering the spinor bi-linears that characterise the
identity structure on $M_5$ \cite{Gauntlett:2005ww}, we find that
all of the above conditions are satisfied if we choose
\begin{equation}
\begin{aligned} \label{Asol}
%&& \label{Asol1} \alpha_0 = e^{2\Delta} Z =0 \\
&  \alpha_0  =0 , \\
&  \alpha_1 = \ft{1}{3} e^{2\Delta} K_3  , \\
&  \beta_2 = \ft{1}{3} e^{4\Delta} V  ,  \\
&  \beta_3 = \ft{1}{3} e^{4\Delta} *_5 V \; .
\end{aligned}
\end{equation}
%\footnote{It is curious to note how $\alpha_0$
%can be taken to be zero. The scalars $\sin \zeta$ and and $S$ do not
%exhaust {\it a priori} all the possibilities for scalar spinor
%bi-linears that one could define on $M_5$ and, indeed, some other
%scalars should in principle  be considered \cite{Gauntlett:2005ww}.
%Among those, there is a complex scalar bi-linear $Z$, constrained by
%supersymmetry to obey the equation $D Z + PZ^* =0$. The same
%equation should be satisfied by $\alpha_0$ (see (\ref{Asystem})),
%which can therefore be assumed to be $\alpha_0 =Z$. However, further
%analysis shows \cite{Gauntlett:2005ww} that only the scalars $\sin
%\zeta$ and $S$ turn out to be independent and, in fact, $Z=0$. This
%might point out towards the existence of deeper reasons that select,
%among all the bilinears and beyond the requirement that it must
%satisfy (\ref{Asystem}), the particular choice (\ref{Asol}) as the
%right one to define the KK ansatz, .}.
The most convenient way to prove this is to again use the specific frame on $M_5$ introduced in
Appendix B of \cite{Gauntlett:2005ww}.
%Moreover, substitution of
%(\ref{Asol}) into (\ref{AEin1}), (\ref{AEin2}), (\ref{AEin3}) also
%leads to constant values of $k_1$, $k_2$, $k_3$ which are,
%precisely, those corresponding to the $D=5$ field equations
%(\ref{D=5Einstein}), (\ref{D=5eomF}). This shows the consitency of
%the truncation of the bosonic sector.

\section{IIB reduction: fermions} \label{AppKKansfer}

Now we show that the KK ansatz (\ref{KKmetric}), (\ref{KKansatzG})
is also consistent at the level of the supersymmetry variations of
the fermions. For this we follow the spinor conventions of Appendix A of
\cite{Gauntlett:2005ww} which we refer the reader to for more details
(we will correct a typo in \cite{Gauntlett:2005ww} below).

The undeformed $AdS_5\times_w M_5$ solution admits Killing spinors of the form
\begin{equation} \label{AKKansepu}
\epsilon = \psi \otimes e^{\Delta /2} \xi_1 \otimes \theta + \psi^c
\otimes e^{\Delta /2} \xi_2^c \otimes \theta ,
\end{equation}
where $\psi$ is a Killing spinor on $AdS_5$, $\theta$ is a constant two-component spinor and,
most importantly, $\xi_1$, $\xi_2$ are Spin(5) spinors on $M_5$ that satisfy
two differential conditions
\begin{align}
   D_m \xi_1
      + \frac{i}{4} \left(e^{-4\Delta}f-2\right)\gamma_m \xi_1
      + \frac{1}{8} e^{-2\Delta} G_{mnp}\gamma^{np}\xi_2
      &= 0 \nn
   \bar{D}_m \xi_2
      -  \frac{i}{4} \left(e^{-4\Delta}f+2\right)\gamma_m \xi_2
      + \frac{1}{8} e^{-2\Delta} G_{mnp}^*\gamma^{np}\xi_1
      &= 0 \label{stwo}
\end{align}
and four algebraic conditions
\begin{align}
   \gamma^m\de_m\Delta\xi_1
      - \frac{1}{48}e^{-2\Delta}\gamma^{mnp}G_{mnp}\xi_2
      - \frac{i}{4}\left(e^{-4\Delta}f-4\right) \xi_1
      &= 0  \nn
   \gamma^m\de_m\Delta\xi_2
      - \frac{1}{48}e^{-2\Delta}\gamma^{mnp}G_{mnp}^*\xi_1
      + \frac{i}{4}\left(e^{-4\Delta}f+4\right)\xi_2
      &= 0 \nn
   \gamma^m P_m \xi_2
      + \frac{1}{24} e^{-2\Delta} \gamma^{mnp} G_{mnp} \xi_1
      &= 0 \nn
   \gamma^m P_m^* \xi_1
      + \frac{1}{24} e^{-2\Delta} \gamma^{mnp} G_{mnp}^* \xi_2
      &= 0 \label{ssix}
\end{align}
where $\gamma^m$ generate Cliff(5) with $\gamma_{12345}=+1$. Note
that $\psi^c = C_{1,4} \psi^*$, $\xi_i^c = C_5 \xi_i^*$, $i=1,2$,
where $C_{1,4}$, $C_5$ are charge conjugation matrices.

The KK ansatz for the $D=10$ Killing spinor is then simply
\begin{equation} \label{AKKanseps}
\epsilon = \varepsilon \otimes e^{\Delta /2} \xi_1 \otimes \theta +
\varepsilon^c \otimes e^{\Delta /2} \xi_2^c \otimes \theta .
\end{equation}
Here $\varepsilon$ is an arbitrary $D=5$ spacetime spinor and the rest is as
in the undeformed case.
For the gravitino, we shall only need a KK reduction
ansatz for the external components, namely (in tangent space):
\begin{equation} \label{AKKanspsi}
\Psi_\alpha = \psi_\alpha \otimes e^{-\Delta /2} \xi_1 \otimes
\theta + \psi_\alpha^c \otimes e^{-\Delta /2} \xi_2^c \otimes \theta
\end{equation}
where $\psi_\alpha$ is the $D=5$ gravitino.

We now demand that the conditions for the KK ansatz to preserve supersymmetry, namely,
that the supersymmetry variations of $\lambda$ and $\Psi_M$
vanish, is the same as the conditions for preservation of supersymmetry in
the $D=5$ gauged supergravity. We will use (\ref{AKKansatz}) but with
$\alpha_0=0$ and $\beta_3 = *_5 \beta_2$.

First consider the variations of the dilatino and of the internal components $\Psi_a$ of the gravitino.
After substituting (\ref{AKKansatz}) into (\ref{susylambda}), (\ref{susypsi})
and using \reef{stwo}, \reef{ssix}, we find that these variations vanish providing that
\begin{equation}
\begin{aligned}
&  \alpha_{1 a} \gamma^a \xi_1 = 0 \\
& \alpha_{1 a}^* \gamma^a \xi_2 = 0 \\[8pt]
& -4h \delta_{a1} \xi_1 +2i e^{-4\Delta} \beta_{2ab} \gamma^b \xi_1
-i e^{-4\Delta} \gamma_{abc} \beta_2^{bc} \xi_1 -e^{-2\Delta}
\alpha_1^b \gamma_{ab} \xi_2
  + 3e^{-2\Delta} \alpha_{1a} \xi_2 =0 \\[8pt]
& -4h \delta_{a1} \xi_2 -2i e^{-4\Delta} \beta_{2ab} \gamma^b \xi_2
+i e^{-4\Delta} \gamma_{abc} \beta_2^{bc} \xi_2 -e^{-2\Delta}
\alpha_1^{*b} \gamma_{ab} \xi_1     + 3e^{-2\Delta} \alpha_{1a}^*
\xi_1 =0 .
\end{aligned}
\end{equation}
One can check that these relations are indeed satisfied\footnote{This can be seen by
using the basis of Cliff(5) and the frame for $M_5$ given in appendix B of \cite{Gauntlett:2005ww}.}
 given our expressions (\ref{Asol}) for $\alpha_1$ and $\beta_2$.

Next consider the variation of the external components of the gravitino.
After substituting (\ref{AKKansatz}) into
(\ref{susypsi}), one finds
\begin{eqnarray}
\delta \Psi_\alpha &=& \ft12 e^{-\Delta /2} \rho_\alpha \varepsilon
\otimes \left( -\ft14 (e^{-4\Delta} f -4) \xi_1 -i \gamma_a
\partial^a \Delta \xi_1 + \ft{i}{48} e^{-2\Delta} \gamma^{abc}
G_{abc} \xi_2 \right) \otimes \theta \nonumber \\[10pt]
&+&  \ft12 e^{-\Delta /2} \rho_\alpha \varepsilon^c \otimes \left(
-\ft14 (e^{-4\Delta} f +4) \xi_2^c -i \gamma_a
\partial^a \Delta \xi_2^c + \ft{i}{48} e^{-2\Delta} \gamma^{abc}
G_{abc} \xi_1^c \right) \otimes \theta \nonumber \\[10pt]
&+ &  e^{-\Delta /2} \Big[  D_\alpha \varepsilon \otimes \xi_1 -
\ft12 \rho_\alpha \varepsilon \otimes \xi_1 - A_\alpha \varepsilon
\otimes \partial_\psi \xi_1 \nonumber \\
&& \qquad +  \ft1{16} F_{\alpha \beta} \rho^\beta \varepsilon
\otimes \left( -4ih \gamma_1 \xi_1 -e^{-4\Delta} \beta_{2ab}
\gamma^{ab} \xi_1  -3i e^{-2\Delta} \alpha_{1a} \gamma^a \xi_2
\right) \nonumber
\\
&& \qquad + \ft1{32} \rho_{\alpha \beta \gamma} F^{\beta \gamma}
\varepsilon \otimes \left( ie^{-2\Delta} \alpha_{1a} \gamma^a \xi_2
 + e^{-4\Delta} \beta_{2bc}
\gamma^{bc} \xi_1 \right) \Big] \otimes \theta \nonumber
\\[10pt]
&+ &  e^{-\Delta /2} \Big[  D_\alpha \varepsilon^c \otimes \xi_2^c +
\ft12 \rho_\alpha \varepsilon^c \otimes \xi_2^c - A_\alpha
\varepsilon^c
\otimes \partial_\psi \xi_2^c \nonumber \\
&& \qquad +  \ft1{16} F_{\alpha \beta} \rho^\beta \varepsilon^c
\otimes \left( -4ih \gamma_1 \xi_2^c -e^{-4\Delta} \beta_{2ab}
\gamma^{ab} \xi_2^c  -3i e^{-2\Delta} \alpha_{1a} \gamma^a \xi_1^c
\right) \nonumber
\\
&& \qquad + \ft1{32} \rho_{\alpha \beta \gamma} F^{\beta \gamma}
\varepsilon^c \otimes \left( ie^{-2\Delta} \alpha_{1a} \gamma^a
\xi_1^c
 + e^{-4\Delta} \beta_{2bc}
\gamma^{bc} \xi_2^c \right) \Big] \otimes \theta \label{psicalc}
\end{eqnarray}
where we are using the coordinate $\psi$ so that the Killing vector
dual to $K_5$ is $3\partial_\psi$. In this expression the
$\rho^\alpha$ generate Cliff(4,1) and satisfy $\rho_{01234}=-i$
(this corrects a sign in \cite{Gauntlett:2005ww}). We also have
$\epsilon_{01234}=+1$.

We now observe that for the choice of forms given in (\ref{Asol})
one has
\begin{eqnarray}
&& -4ih \gamma_1 \xi_1 -e^{-4\Delta} \beta_{2ab} \gamma^{ab} \xi_1
-3i e^{-2\Delta} \alpha_{1a}  \gamma^a \xi_2 = -\ft{8i}{3} \xi_1
\nonumber \\[10pt]
&& i e^{-2\Delta} \alpha_{1a} \gamma^a \xi_2 + e^{-4\Delta}
\beta_{2bc} \gamma^{bc} \xi_1 = \ft{4i}{3} \xi_1
\end{eqnarray}
and similar expressions for the last two terms of (\ref{psicalc}).
Using these results, the fact that $\partial_\psi \xi_1 = -
\ft{i}{2} \xi_1$ and equations \reef{ssix}, after introducing the KK
ansatz (\ref{AKKanspsi}) for the gravitino we deduce that
\begin{eqnarray}
&& \delta \psi_\alpha \otimes e^{-\Delta /2} \xi_1 \otimes \theta  +
\delta \psi_\alpha^c \otimes e^{-\Delta /2} \xi_2^c \otimes \theta = \nonumber \\[10pt]
&& \qquad  \left( D_\alpha \varepsilon -\ft12\rho_\alpha \varepsilon
+ \ft{i}{2} A_\alpha \varepsilon + \ft{i}{24} F_{\beta \gamma}
(\rho_\alpha{}^{\beta \gamma} -4 \delta_\alpha^\beta \rho^\gamma )
\varepsilon  \right) \otimes \ e^{-\Delta/2} \xi_1
\otimes \theta \nonumber \\[10pt]
&& \qquad  \left( D_\alpha \varepsilon^c +\ft12 \rho_\alpha
\varepsilon^c - \ft{i}{2} A_\alpha \varepsilon^c + \ft{i}{24}
F_{\beta \gamma} (\rho_\alpha{}^{\beta \gamma} -4
\delta_\alpha^\beta \rho^\gamma ) \varepsilon^c  \right) \otimes \
e^{-\Delta/2} \xi_2^c \otimes \theta ,\nonumber \\ &&
\end{eqnarray}
which implies
\begin{equation}
\delta \psi_\alpha = D_\alpha \varepsilon -\ft12 \rho_\alpha
\varepsilon + \ft{i}{2} A_\alpha \varepsilon + \ft{i}{24} F_{\beta
\gamma} (\rho_\alpha{}^{\beta \gamma} -4 \delta_\alpha^\beta
\rho^\gamma ) \varepsilon ,
\end{equation}
as claimed in the text.

\end{document}